\documentclass[prb,twocolumn,preprintnumbers]{revtex4}

\usepackage{graphicx}
\usepackage{dcolumn}
\usepackage{amsmath}
\usepackage{color}

\begin{document}

\title{Metal-insulator transition and local-moment collapse in negative charge-transfer CaFeO$_3$ under pressure}

\author{I. Leonov}
\affiliation{M. N. Miheev Institute of Metal Physics, Russian Academy of Sciencee
s, 620108 Yekaterinburg, Russia}
\affiliation{Ural Federal University, 620002 Yekaterinburg, Russia}

\begin{abstract}
We compute the electronic structure, spin and charge state of Fe ions, and structural phase stability of paramagnetic CaFeO$_3$ under pressure using a fully self-consistent in charge density DFT+dynamical mean-field theory method.
We show that at ambient pressure CaFeO$_3$ is a negative charge-transfer insulator characterized by strong localization of the Fe $3d$ electrons. It crystallizes in the monoclinic $P2_1/n$ crystal structure with a cooperative breathing mode distortion of the lattice. 
While the Fe $3d$ Wannier occupations and local moments are consistent with robust charge disproportionation of Fe ions in the insulating $P2_1/n$ phase, the physical charge density difference around the structurally distinct Fe A and Fe B ions with the ``contracted'' and ``expanded'' oxygen octahedra, respectively, is rather weak, $\sim$0.04. This implies the importance of the Fe $3d$ and O $2p$ negative charge transfer and supports the formation of a bond-disproportionated state characterized by the Fe A $3d^{5-\delta}\underline{L}^{2-\delta}$ and Fe B $3d^5$ valence configurations with $\delta \ll 1$, in agreement with strong  hybridization between the Fe $3d$ and O $2p$ states.
Upon compression above $\sim$41 GPa CaFeO$_3$ undergoes the insulator-to-metal phase transition (IMT) which is accompanied by a structural transformation into the orthorhombic $Pbnm$ phase.
The phase transition is accompanied by suppression of the cooperative breathing mode distortion of the lattice and, hence, results in the melting of bond disproportionation of the Fe ions. Our analysis suggests that the IMT transition is associated with orbital-dependent delocalization of the Fe $3d$ electrons and leads to a remarkable collapse of the local magnetic moments. Our results imply the crucial importance of the interplay of electronic correlations and structural effects to explain the properties of CaFeO$_3$. 

\end{abstract}

\maketitle

%%%%%%%%%%%%%%%%%%%%%%%
\section{Introduction}
%%%%%%%%%%%%%%%%%%%%%%%

The transition metal oxides with perovskite crystal structure have attracted much attention due to their diverse electronic and magnetic properties, allowing for a broad range of applications, e.g., in electronics and spintronics, and energy storage \cite{Mott_1990,Imada_1998,Tokura_2000,Dagotto_2001,Salamon_2001,Tokura_2014}. A particular interest is given to the understanding of a Mott metal-insulator transition (MIT) driven by correlation effects, the description of which has been a long-standing challenge \cite{Mott_1990,Imada_1998}.
In this context metal oxides with an unusually high valence state of the transition metal, Ni$^{3+}$ (nickelates) and Fe$^{4+}$ (ferrates) are notable because of a complex interplay of ligand holes and electronic correlations in the $3d$ shell of Ni and Fe ions \cite{Catalan_2008,AnnuRevMaterRes.46.305,RepProgPhys.81.46501,Takano_1991,Bocquet_1992,Woodward_2000,Takeda_2000}, which leads to their complex physical behavior. 

Such materials are often regarded as ``negative charge transfer'' compounds, implying the importance for a charge transfer from oxygen to the transition metal atom, leaving holes at oxygen ligands, to explain their electronic properties \cite{Zaanen_1985,Mazin_2007}. In particular, the rare-earth nickelate perovskites $R$NiO$_3$ ($R$ = rare earth, $R^{3+}$) with a high oxidation state of nickel, Ni$^{3+}$ $3d^7$ exhibit a sharp metal-insulator transition (except for LaNiO$_3$) upon cooling and under pressure \cite{Catalan_2008,AnnuRevMaterRes.46.305,RepProgPhys.81.46501,Torrance_1992,Garcia_Munoz_1992}. The MIT transition is accompanied by a structural transformation from the orthorhombic metallic ($Pbnm$, distorted GdFeO$_3$-type) to monoclinic insulating ($P2_1/n$) crystal structure, with a cooperative breathing mode distortion of NiO$_6$ octahedra in the insulating $P2_1/n$ phase \cite{Torrance_1992,Garcia_Munoz_1992}. While analysis of the Ni-O bond lengths and x-ray absorption spectroscopy suggests charge disproportionation of Ni ions in the insulating $R$NiO$_3$ phases, their electronic state is more accurately described in terms of bond disproportionation, with alternating Ni ions which (nearly) adopt a Ni$^{2+}$ $3d^8$ (Ni$^{2+}$ ions with local moments) and $3d^8\underline{L}^2$ (nonmagnetic spin-singlet) electronic configuration ($\underline{L}$ denotes a hole in the O $2p$ band) \cite{Park_2012,Johnston_2014,Park_2014,Subedi_2015,Bisogni_2016,Hampel_2017,Hampel_2019,Georgescu_2019,Peil_2019}. We also note that the complex interplay between (negative) charge transfer and correlation effects can result in unusual charge or bond-disproportionation of the $A$-site ions of the perovskite ABO$_3$ structure, e.g., in BiNiO$_3$ \cite{Ishiwata_2002,Azuma_2007,Carlsson_2008,Azuma_2011,Oka_2013,Saha_Dasgupta_2019,Leonov_2020b}, PbFeO$_3$ \cite{Tsuchiya_2006,Ye_2021}, and PbCoO$_3$ \cite{Sakai_2017,Liu_2020,Hariki_2021}.

In the present work, we focus on a negative charge transfer material CaFeO$_3$ with a nominal Fe $3d^4$ $(t_{2g}^3e_g^1)$ electronic state that exhibits the MIT upon cooling below $\sim$290 K (at ambient pressure) \cite{Catalan_2008,AnnuRevMaterRes.46.305,RepProgPhys.81.46501,Woodward_2000,Takeda_2000}. The MIT occurs upon compression of CaFeO$_3$ above $\sim$30 GPa \cite{Takano_1991}. It shows a complex screw antiferromagnetic structure below the N\'eel temperature $\sim$115 K \cite{Woodward_2000,Takeda_2000}. Similarly to the rare-earth nickelates, the MIT in CaFeO$_3$ is accompanied by a structural transformation from the orthorhombic metallic ($Pbnm$) to monoclinic insulating ($P2_1/n$) crystal structure, with a cooperative breathing mode distortion of FeO$_6$ octahedra with the difference in Fe-O bond lengths $\sim$0.1\AA\ in the insulating $P2_1/n$ phase \cite{Woodward_2000,Takeda_2000}. 

Given the electronic and structural complexity of CaFeO$_3$ (e.g., doped with Ca and La) and its similarity to the rare-earth nickelates, this system has been intensively studied using both experimental and theoretical methods \cite{Bocquet_1992,Woodward_2000,Takeda_2000,Takano_1991,Xu_2001,Patrakeev_2003,Galakhov_2010,Fujioka_2012,Rogge_2018,Rogge_2019,Jana_2019,Wang_2019,Onose_2020,Shein_2005,Saha-Dasgupta_2005,Alexandrov_2008,Dalpian_2018,Cammarata_2012,Zhang_2018,Varignon_2019,Varignon_2019b}. Previous calculations were mostly employed band-structure methods supplemented with the on-site Coulomb correlations for the Fe $3d$ states within density-functional theory (DFT)+$U$ method \cite{Anisimov_1991,Liechtenstein_1995} with a major focus to the ground state properties of the insulating $P2_1/n$ phase with a long-range magnetic ordering \cite{Shein_2005,Saha-Dasgupta_2005,Alexandrov_2008,Cammarata_2012,Zhang_2018}. Using the DFT+$U$ and more advanced strongly constrained appropriately normed exchange and correlation functional (SCAN) methods \cite{Anisimov_1991,Liechtenstein_1995,Sun_2015} the authors address the questions about the origin of a band gap, charge and bond-disproportionation, ligand hole effects, Mott vs. charge transfer insulator behaviors, etc. in CaFeO$_3$ \cite{Dalpian_2018,Varignon_2019,Varignon_2019b}.

Recently, the electronic structure of CaFeO$_3$ has been discussed \cite{Merkel_2021} in the context of a five-orbital tight-binding model using the DFT+dynamical mean-field theory (DFT+DMFT) method \cite{Metzner_1989,Georges_1996,Kotliar_2006,Anisimov_1997,Haule_2007,Pourovskii_2007,Amadon_2008,Aichhorn_2009}. 
DFT+DMFT has been proven to be among the most advanced theoretical methods for studying the electronic properties of strongly correlated materials, such as correlated transition metal oxides, heavy-fermions, and Fe-based superconductors, e.g., to study the Mott metal-insulator phase transition, collapse of local moments, large orbital-dependent renormalizations, etc. \cite{Haule_2007,Pourovskii_2007,Amadon_2008,Aichhorn_2009,Leonov_2015a,Greenberg_2018,Leonov_2019,Leonov_2015b,Leonov_2015c,Zhong_2015,Leonov_2016,Backes_2016,Medici_2017,Seth_2017,Facio_2018,Skornyakov_2017,Lechermann_2018,Arribi_2018,Skornyakov_2018,Lechermann_2019,Jang_2019,Mandal_2019,Leonov_2020,Koemets_2021,Leonov_2021,Leonov_2021b,Lantz_2015,Bhandary_2021}
The DFT+DMFT results for the tight-binding model \cite{Merkel_2021} show the competition between high-spin and low-spin homogeneous and an inhomogeneous charge-disproportionated state, implying the importance of Hund's coupling \cite{Georges_2013,Kunes_2011,Isidori_2019}.
Nonetheless, the electronic properties of CaFeO$_3$ and, in particular, the interplay of the electronic structure, magnetic states, and phase stability of CaFeO$_3$ near the MIT are still poorly understood. 

We address this problem in our study by using a fully self-consistent in charge density DFT+DMFT method \cite{Haule_2007,Pourovskii_2007,Amadon_2008,Aichhorn_2009,Leonov_2015a,Leonov_2015c,Leonov_2020} implemented with plane-wave pseudopotentials \cite{Giannozzi_2009}. By employing DFT+DMFT it becomes possible to capture all generic aspects of the interplay between the electronic correlations, magnetic states, and crystal structure of CaFeO$_3$ under pressure. In particular, we explore the electronic structure, magnetic properties, spin and  charge state of Fe ions, and structural phase stability of \emph{paramagnetic} CaFeO$_3$ near the pressure-induced MIT.
Our results show the importance of the Fe $3d$ and O $2p$ negative charge transfer due to strong covalency and supports the formation of the bond-disproportionated state in the insulating phase of CaFeO$_3$

%%%%%%%%%%%%%%%%%
\section{Method}
%%%%%%%%%%%%%%%%%

We employ the state-of-the-art DFT+DMFT method \cite{Georges_1996,Kotliar_2006,Anisimov_1997} with full self-consistency over the charge density \cite{Haule_2007,Pourovskii_2007,Amadon_2008,Aichhorn_2009,Leonov_2015a,Leonov_2015c,Leonov_2020} to explore the electronic structure, spin and charge state of Fe ions, and lattice properties of paramagnetic CaFeO$_3$ under pressure.
We start by constructing the effective low-energy Hamiltonian for the partially occupied Fe $3d$ and O $2p$ states and consider the multiorbital Hubbard Hamiltonian with the local electron-electron interaction part (in the density-density approximation) for the Fe $3d$ orbitals
\begin{eqnarray}
\label{eq:hamilt}
\hat{H}_\mathrm{int} &=& \frac{1}{2} \sum_{mm'\sigma} U_{mm'} \hat{n}_{m\sigma} \hat{n}_{m'\bar{\sigma}} \nonumber \\ 
&+& \frac{1}{2} \sum_{m \neq m'\sigma} (U_{mm'} - J_{mm'}) \hat{n}_{m\sigma} \hat{n}_{m'\sigma},
\end{eqnarray}
where $\hat{n}_{m\sigma}$ is the occupation number operator with spin $\sigma$ and (diagonal) orbital indices $m$. $U_{mm'}$ and $J_{mm'}$ are the Coulomb repulsion and Hund's exchange coupling matrix elements. In Eq.~\ref{eq:hamilt} we assume summation over all Fe sites.
For the partially filled Fe $3d$ and O $2p$ orbitals we construct a basis set of atomic-centered symmetry-constrained Wannier functions \cite{Marzari_2012,Anisimov_2005,Trimarchi_2008} (the localized Fe $3d$ orbitals are constructed using the Fe $3d$ band set, while the O $2p$ orbitals were defined over the full energy window spanned by the O $2p$–Fe $3d$ band complex \cite{Greenberg_2018,Leonov_2019}).
This makes it possible to explicitly account charge transfer effects between the partially occupied Fe $3d$ and O $2p$ states and strong Coulomb correlations of the Fe $3d$ electrons. We use the continuous-time hybridization expansion (segment) quantum Monte Carlo algorithm in order to solve the realistic many-body problem \cite{Gull_2011}. 
The DFT+DMFT calculations are performed in the paramagnetic (PM) state at an electronic temperature $T=387$ K.
The Coulomb $U_{mm'}$ and Hund's exchange $J_{mm'}$ matrix elements are parametrized in terms of the Slater integrals $F_0$, $F_2$, and $F_4$ which are  expressed in terms of the average Coulomb repulsion $U = F_0$ and Hund’s exchange coupling $J_H = (F_2 + F_4)/14$, with a fixed ratio $F_4/F_2 = 0.63$. We take the average Hubbard $U = 6$~eV and Hund's exchange $J_H = 0.86$~eV in accordance with previous DFT+DMFT studies \cite{Greenberg_2018,Leonov_2019}. The fully localized double-counting correction evaluated from the self-consistently determined local occupations was used, 
$\hat{H}_{DC}=U ( N - \frac{1}{2} ) - J ( N_{\sigma} - \frac{1}{2} )$,
where $N_\sigma$ is the total Fe $3d$ occupation with spin $\sigma$ and $N=N_\uparrow+N_\downarrow$.
The Coulomb interaction is treated in the density-density approximation, neglecting by spin-flip and pair-hopping terms in the multiorbital Hubbard Hamiltonian. The spin-orbit coupling is neglected in our calculations. The spectral functions were computed using the maximum entropy method. 
In order to estimate the quasiparticle mass enhancement $m^*/m$ of the Fe $3d$ states we perform analytic continuation of the self-energy $\Sigma(\omega)$ determined self-consistently on the Matsubara contour using Pad\'e approximants. In order to analyse a degree of localization of the Fe $3d$ electrons we compute the local spin susceptibility $\chi(\tau)=\langle \hat{m}_z(\tau)\hat{m}_z(0)\rangle$ within DMFT, where $\hat{m}_z(\tau)$ is the instantaneous magnetization on the Fe $3d$ site at the imaginary time $\tau$ ($\tau$ denotes an imaginary-time evolution ranging from 0 to $\beta=1/k_BT$ in the path integral formalism).
In DFT we use the generalised gradient Perdew-Burke-Ernzerhof approximation for the correlation exchange functional \cite{Perdew_1996}. 

%%%%%%%%%%%%%%%%%%%%%
\section{Results and discussion}
%%%%%%%%%%%%%%%%%%%%%

We start by computing the electronic properties and structural phase stability of PM CaFeO$_3$ under pressure. In our study, we adopt the crystal structure parameters determined experimentally for the monoclinic $P2_1/n$ phase (at ambient pressure and temperature $\sim$15~K) and orthorhombic $Pbnm$ structures (taken above the insulator-to-metal and charge-disproportionation transition, at 300~K) \cite{Woodward_2000,Takeda_2000}, and evaluate the DFT+DMFT total energies as a function of the unit-cell volume \cite{Leonov_2015a,Leonov_2015c,Leonov_2020}. We fit the total-energy results using the third-order Birch-Murnaghan equation of states separately for the low- and high-volume regions \cite{Murnaghan_1944,Birch_1947}. 

\begin{figure}[tbp!]
\centerline{\includegraphics[width=0.4\textwidth,clip=true]{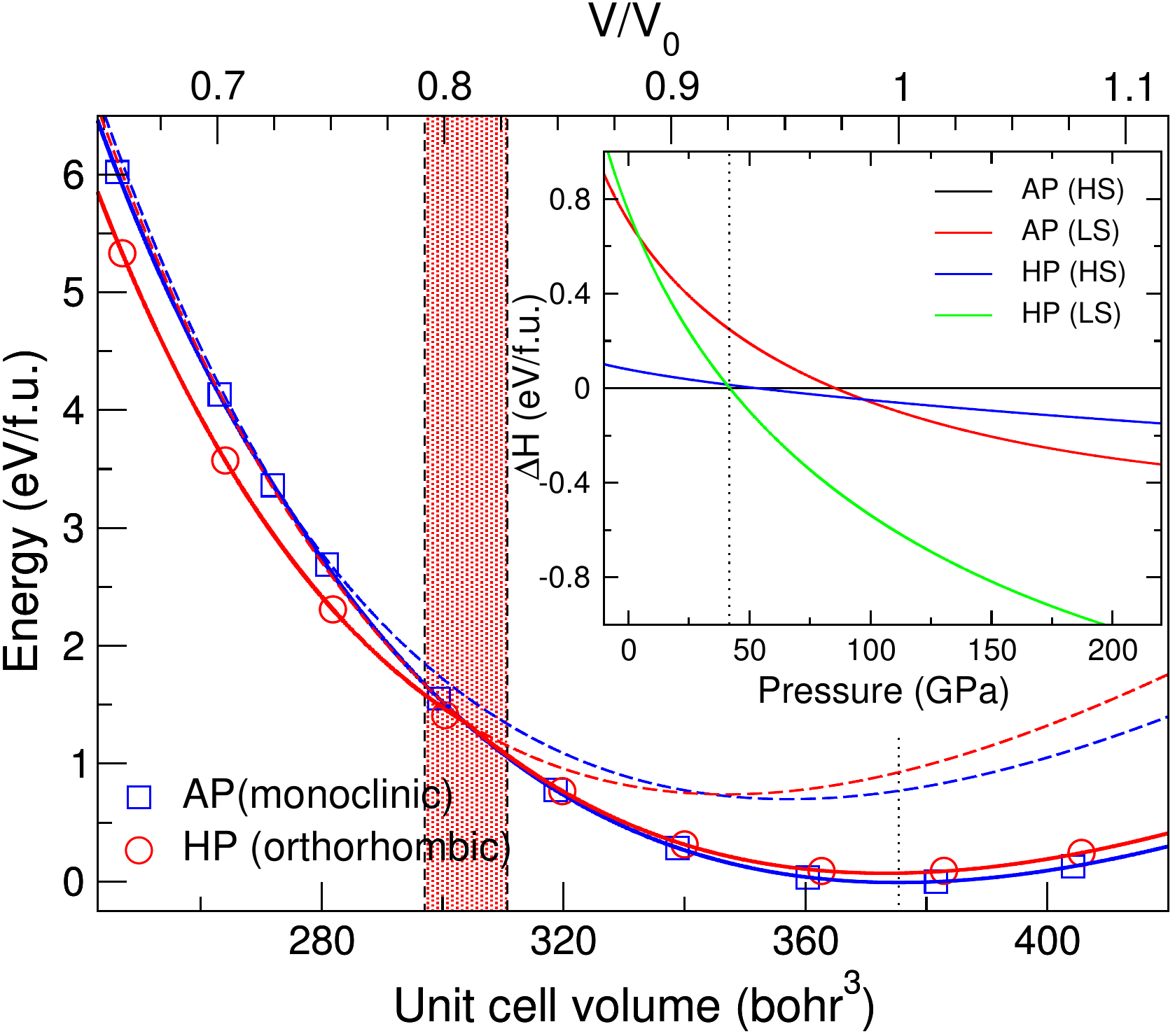}}
\caption{Total energy of PM CaFeO$_3$ calculated by DFT+DMFT at $T=387$~K as a function of lattice volume. The lattice volume collapse associated with the $P2_1/n$-$Pbnm$ structural phase transition is depicted by red shading. Inset: Enthalpy difference (relative to the high-spin $P2_1/n$ solution) obtained by DFT+DMFT as a function of pressure. The $P2_1/n$-$Pbnm$ structural transition at $\sim$41 GPa is shown by a vertical dotted line.
}
\label{Fig_1}
\end{figure}

\begin{figure}[tbp!]
\centerline{\includegraphics[width=0.4\textwidth,clip=true]{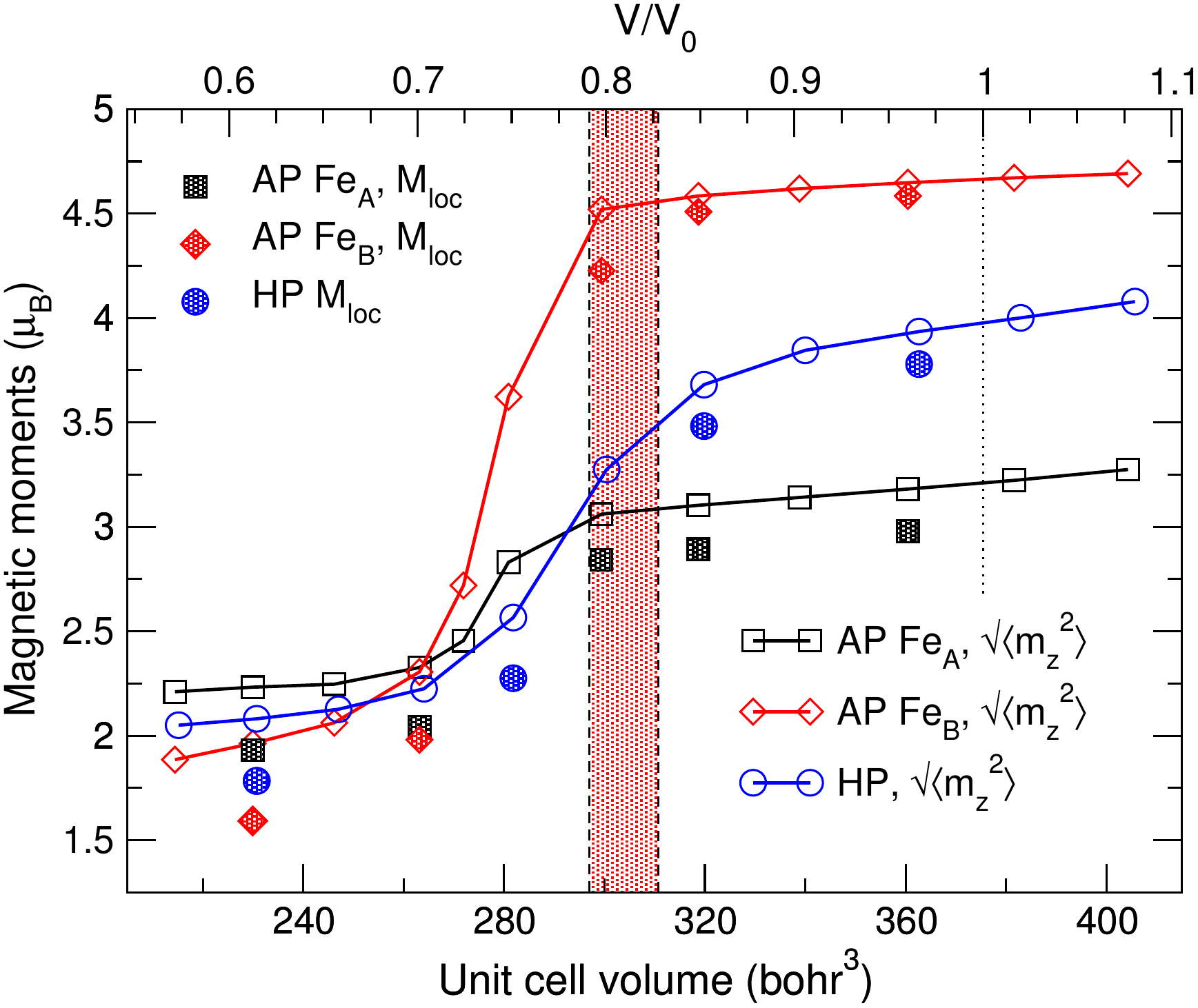}}
\caption{The local magnetic moments of the $P2_1/n$ (AP) and $Pbnm$ (HP) structural phases of PM CaFeO$_3$ obtained by DFT+DMFT at $T=387$~K as a function of the unit-cell volume. The unit-cell volume collapse associated with the MIT and the $P2_1/n$-$Pbnm$ structural transition is depicted by red shading.
}
\label{Fig_2}
\end{figure}

Our results for the total energy calculations and compressional behavior of CaFeO$_3$ are summarized in Fig.~\ref{Fig_1}. The $P2_1/n$ phase is found to be energetically favorable, i.e., thermodynamically stable at ambient pressure, with a total-energy difference of $\sim$13 meV/atom with respect to the $Pbnm$ phase. This implies that at low pressure and temperature conditions CaFeO$_3$ crystallizes in the $P2_1/n$ phase which is characterized by the cooperative breathing-mode distortion of the crystal structure, in agreement with the x-ray and neutron diffraction experiments \cite{Woodward_2000,Takeda_2000}. Given the higher electronic entropy of the metallic $Pbnm$ phase this also suggests a structural phase transition from $P2_1/n$ to the $Pbnm$ phase upon heating, consistent with experiments \cite{Woodward_2000,Takeda_2000}.
%Moreover, our results suggest a structural phase transition from $P2_1/n$ to the $Pbnm$ phase upon heating caused by a higher electronic and lattice entropy of the metallic $Pbnm$ phase. 
The calculated equilibrium lattice volume for the $P2_1/n$ phase is $V_0 = 375.4$ a.u.$^3$ (by $\sim$4\% larger than that in the experiment) and bulk modulus $K_0=151$ GPa (with $K'\equiv dK/dP$ fixed to $K'=4.0$). 
Our results for the instantaneous magnetic moments of the structurally distinct Fe A and Fe B sites with the ``contracted'' and ``expanded'' oxygen octahedra, respectively, are 3.18 and 4.65$\mu_\mathrm{B}$ (see Fig.~\ref{Fig_2}).
These values are close to the calculated fluctuating moments $\sim$2.98 and 4.58$\mu_\mathrm{B}$, evaluated as the imaginary-time average of local spin susceptibility $\chi(\tau)=\langle \hat{m}_z(\tau)\hat{m}_z(0)\rangle$, $m_\mathrm{loc}=(k_BT \int \chi(\tau) d\tau)^{1/2}$. This suggests strong localization of the Fe $3d$ electrons at low pressures, implying the crucial importance of electronic correlations to explain the electronic properties of CaFeO$_3$. We also note that the calculated local moments are in good agreement with the experimental estimates of 2.5-3.5$\mu_\mathrm{B}$ and 3.5-5.0$\mu_\mathrm{B}$ for the Fe A and B sites obtained from the fit of neutron diffraction data using different spiral magnetic structures \cite{Woodward_2000,Takeda_2000}. 

\begin{figure}[tbp!]
\centerline{\includegraphics[width=0.5\textwidth,clip=true]{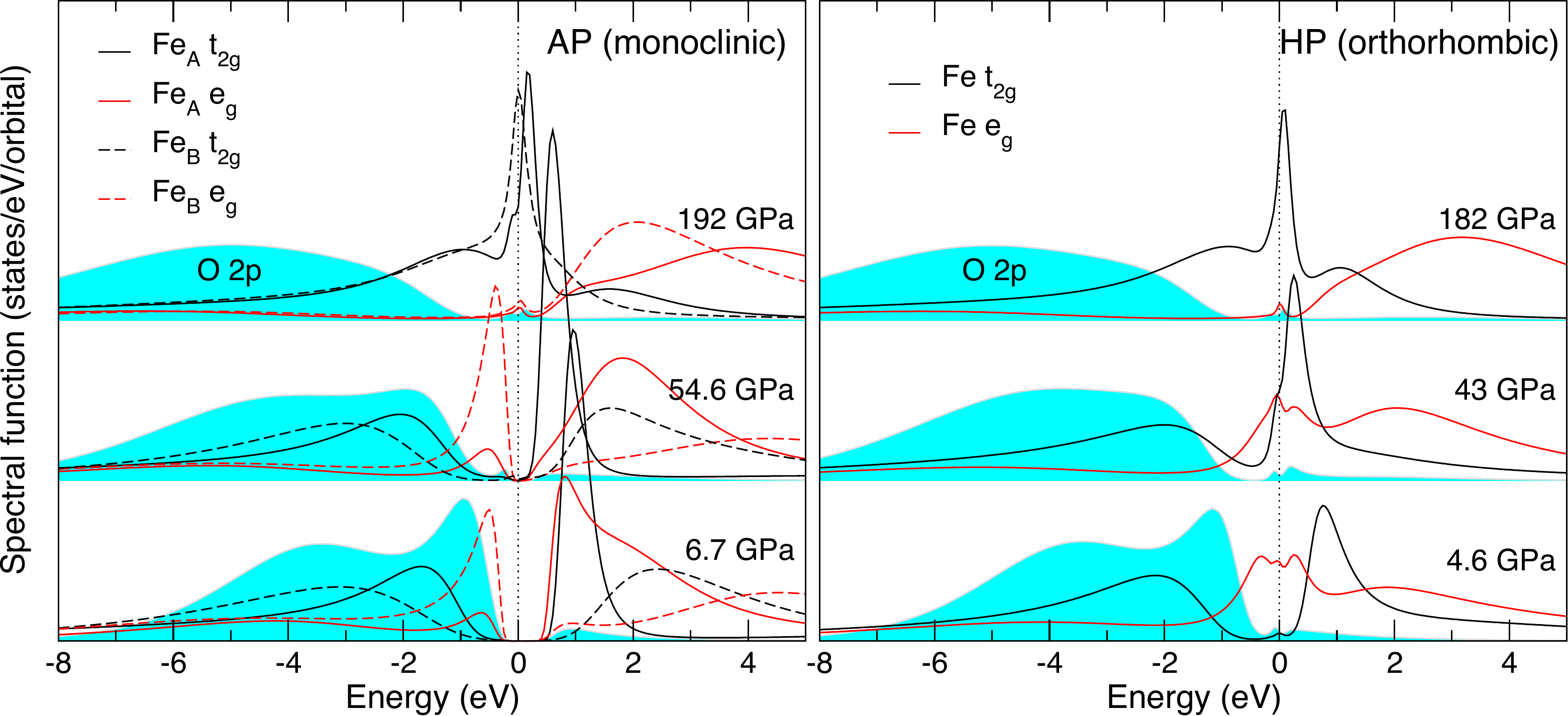}}
\caption{Orbitally-resolved spectral functions of PM CaFeO$_3$ calculated by DFT+DMFT for the $P2_1/n$ (left panel) and $Pbnm$ (right panel) phases for different lattice volumes.
}
\label{Fig_3}
\end{figure}

In agreement with experiment, we obtain a negative charge-transfer insulator \cite{Zaanen_1985,Mazin_2007} with an energy gap of $\sim$0.6~eV (see the left panel of Fig.~\ref{Fig_3}). Moreover, our analysis of the Fe $3d$ Wannier occupations give a remarkable charge disproportionation between the Fe A and Fe B sites (due to a sufficiently different oxygen environment of these sites in the insulating $P2_1/n$ phase, with the difference in Fe-O bond lengths $\sim$0.1~\AA \cite{Woodward_2000,Takeda_2000}). In fact, at $\sim$6.7 GPa the total Fe $3d$ Wannier occupation for the Fe A site is only $\sim$4.3, while for the Fe B sites it is 5.1, implying a rather large charge difference of $\sim$0.8  electron. We note that this charge difference is about 40\% of the ideal ionic Fe$^{3+}$-to-Fe$^{5+}$ charge disproportionation, roughly consistent with the bond valence sum estimate of $\sim$1.1 \cite{Woodward_2000,Takeda_2000}, as well as is in agreement with the significantly different local magnetic moments for the Fe A and Fe B sites (see above). Interestingly that previous estimates of a charge disproportionation in the low-temperature charge-ordered phases of the mixed-valence oxides such as Fe$_3$O$_4$ and rare-earth nickelates $R$NiO$_3$ give $\sim$20-40\% of the ideal ionic disproportionation \cite{Wright_2002,Leonov_2006,Jeng_2006,Leonov_2005,Senn_2012,Perversi_2019,Torrance_1992,Garcia_Munoz_1992,Catalan_2008}, consistent with our results. 

In agreement with this, our analysis of the eigenvalues of the reduced Fe $3d$ density matrix, i.e., an estimate of a fluctuating valence state of Fe ions, shows that the Fe B ions (with ``expanded'' oxygen octahedra) are $3+$. In fact, for the Fe B ions $3d^5$ configuration has a predominant weight of about 74\% with a $\sim$7\% admixture of the $3d^4$ and $\sim$18\% of the $3d^6$ configurations (due to quantum mixing and temperature effects). On the other hand, for the Fe A ions the $3d^4$ and $3d^5$ configurations are nearly equally weighted, of $\sim$48\% and 34\%, respectively, in accordance with above estimates. 
That is, the electronic state of the Fe A ions consists of heavily mixed $d^4$ and $d^5\underline{L}$ states, reflecting  the importance of Fe $3d$ to O $2p$ negative charge transfer due to strong covalency in CaFeO$_3$. 

While our results for the Fe $3d$ Wannier occupations and local moments give a robust charge disproportionation in the $P2_1/n$ insulating phase, a difference of the total charge density $\rho({\bf r})$ around the structurally distinct Fe A and Fe B ions is rather weak. 
In particular, our result for the corresponding charge difference within the Fe-ion radius of 0.86\AA\ give an order of magnitude smaller value of $\sim$0.04. This implies the importance of the Fe $3d$ and O $2p$ negative charge transfer and suggests the formation of a bond-disproportionated state characterized by the Fe $3d^{5-\delta}\underline{L}^{2-\delta}$ and $3d^5$ electronic configurations with $\delta \ll 1$ for the ``compressed'' Fe A and ``expanded'' Fe B sites, respectively, in the insulating phase of CaFeO$_3$. This result is consistent with a substantial Fe-O covalence and strong hybridization between the unoccupied Fe $e_g$ and O $2p$ state \cite{Zaanen_1985,Mazin_2007}. 

Our results for the orbitally-resolved spectral functions of PM CaFeO$_3$ are shown in Fig.~\ref{Fig_3}. The energy gap lies between the occupied Fe B $e_g$ states strongly mixed with the O $2p$ and the unoccupied Fe A $3d$ states. The latter strongly hybridize with the empty O $2p$ states. The occupied O $2p$ band appears at about -6 eV to near the Fermi level. The top of the valence band has a strongly mixed Fe $3d$ and O $2p$ character, with a resonant peak in the filled Fe $e_g$ states located at about -0.4~eV below the Fermi level. This behavior can be ascribed to the formation of a Zhang-Rice bound state \cite{Zhang_1988}. 

Fe $t_{2g}$ and $e_g$ Wannier orbital occupancies are 0.55 and 0.25 per spin-orbit for the Fe A and are 0.52 and 0.5 for the Fe B ions, respectively. These findings clearly indicate that at ambient pressure the Fe B ions are in the high-spin (HS) state.
In fact, in an ionic picture, the Fe$^{3+}$ ions have a $3d^5$ configuration with three electrons in the $t_{2g}$ and two in the $e_g$ orbitals (in the octahedral crystal field), forming the $S=5/2$ state. 
This result is in agreement with our results for the decomposition of the electronic state into atomic spin-state configurations within DFT+DMFT. In particular, for the Fe B ions the HS state has a predominant weight of $\sim$97\% with a small admixture of $\sim$3\% due to the intermediate-spin (IS) state in the insulating $P2_1/n$ phase (at ambient pressure).
On the other hand, the Fe A ions show a strong mixture of the HS and IS state configurations whose corresponding weights are $\sim$47\% and 43\%.

\begin{figure}[tbp!]
\centerline{\includegraphics[width=0.5\textwidth,clip=true]{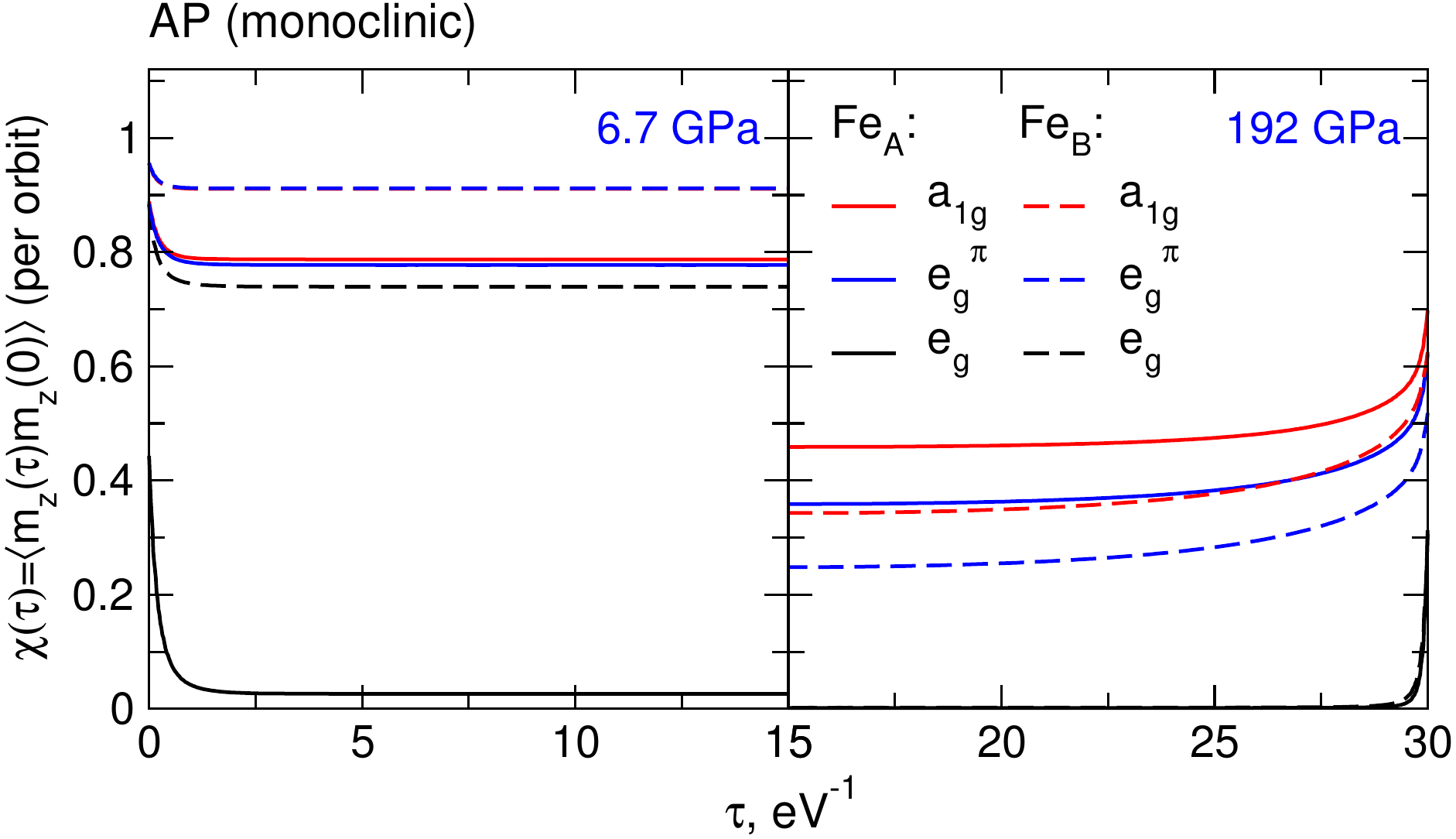}}
\caption{Orbitally-resolved local spin correlations functions $\chi(\tau)=\langle \hat{m}_z(\tau) \hat{m}_z(0) \rangle$ of the $P2_1/n$ phase of CaFeO$_3$ calculated by DFT+DMFT for different pressures.
}
\label{Fig_4}
\end{figure}

Under pressure the energy gap of $P2_1/n$ CaFeO$_3$ gradually decreases, resulting in an insulator-to-metal phase transition, within the $P2_1/n$ structure upon compression to $\sim$0.74~$V_0$, above $\sim$85 GPa. The phase transition is accompanied by a remarkable anomaly of our DFT+DMFT total-energy and local-moment results. In fact, the local moments of the Fe A and B ions are significantly different and are seen to retain their high-spin values down to $\sim$0.74~$V_0$, while upon further compression collapse to a nearly same moment value of $\sim$2.3$\mu_\mathrm{B}$, at 102 GPa. The phase transition is accompanied by a large transfer of the Fe $3d$ spectral weight, with a formation of the Fe $3d$ quasiparticle peak near the Fermi level. It is associated with a substantial redistribution of charge within the Fe $3d$ shell. 
Fe $t_{2g}$ orbital occupations are found to gradually increase with pressure ($t_{2g}$ occupation is about 0.66 per spin-orbit for the Fe A and 0.69 for the Fe B sites at 102 GPa). 
This comes along with a significant depopulation of the Fe $e_g$ states. Fe A and B $e_g$ occupations are 0.17 and 0.19 per spin-orbit, respectively, while the Wannier Fe $3d$ total occupancies change to 4.6 and 4.87 for the Fe A and Fe B ions. This implies a significant reduction of charge disproportionation upon metallization of the $P2_1/n$ structure, to about 0.26 at 102 GPa.

Our results for the local (dynamical) spin-spin correlation function $\chi(\tau)=\langle \hat{m}_z(\tau) \hat{m}_z(0)\rangle$ reveal significant delocalization of the Fe $3d$ electrons under pressure as seen in Fig.~\ref{Fig_4}. $\chi(\tau)$ is strongly suppressed, quickly decaying with the imaginary time $\tau$ (it is most pronounced for the Fe B $e_g$ sates). On the other hand, a relatively large value of $\chi(\tau)$ of $\sim$0.3--0.4 $\mu_B^2$ at $\tau\simeq \beta/2$ (with $\beta=1/k_BT=30$ eV$^{-1}$ at $T=387$ K) for the Fe A and B $t_{2g}$ states suggests the proximity of Fe $3d$ magnetic moments to orbital-selective localization.

However, we note that the calculated critical pressure of the MIT $p_c \sim 85$ GPa in the $P2_1/n$ phase of CaFeO$_3$ is significantly, by almost three times, higher than that found in the experiments \cite{Takano_1991}. It is know experimentally that the MIT in CaFeO$_3$ is accompanied by a structural change to the orthorhombic $Pbnm$ phase \cite{Takano_1991,Bocquet_1992,Woodward_2000,Takeda_2000}. In agreement with this, our DFT+DMFT total-energy calculations show that upon compression to $\sim$0.83~$V_0$, the $P2_1/n$ lattice becomes energetically unfavorable, with the orthorhombic $Pbnm$ phase being thermodynamically stable. We therefore conclude that above $\sim$41 GPa CaFeO$_3$ undergoes a phase transition from monoclinic $P2_1/n$ to the orthorhombic $Pbnm$ crystal structure (see Fig.~\ref{Fig_1}). In agreement with experiment, the phase transition is accompanied by metallization of CaFeO$_3$. Interestingly, the calculated local magnetic moments for the $Pbnm$ phase are nearly equal to that in the highly compressed metallic $P2_1/n$ phase of CaFeO$_3$, $\sim$2.22$\mu_\mathrm{B}$ at $\sim$0.7~$V_0$.
The phase transition is accompanied by a remarkable collapse of the unit-cell volume of $\frac{\Delta V}{V} \sim 4$\% and results in suppression of the cooperative breathing mode distortion of the lattice. In fact, in the $Pbnm$ phase Fe sites have a regular oxygen coordination with no evidence for charge disproportionation. We also note that the charge disproportionated state is found to be unstable within our DFT+DMFT calculations for the $Pbnm$ lattice. The calculated bulk modulus for the high-pressure $Pbnm$ phase is about 197 GPa. This suggests that the $P2_1/n$ to $Pbnm$ phase transition is accompanied by a remarkable \emph{increase} of the bulk modulus upon the pressure-induced insulator-to-metal phase transition, implying a crucial importance of the interplay of electronic correlations and the lattice to explain the properties of CaFeO$_3$.

\begin{figure}[tbp!]
\centerline{\includegraphics[width=0.5\textwidth,clip=true]{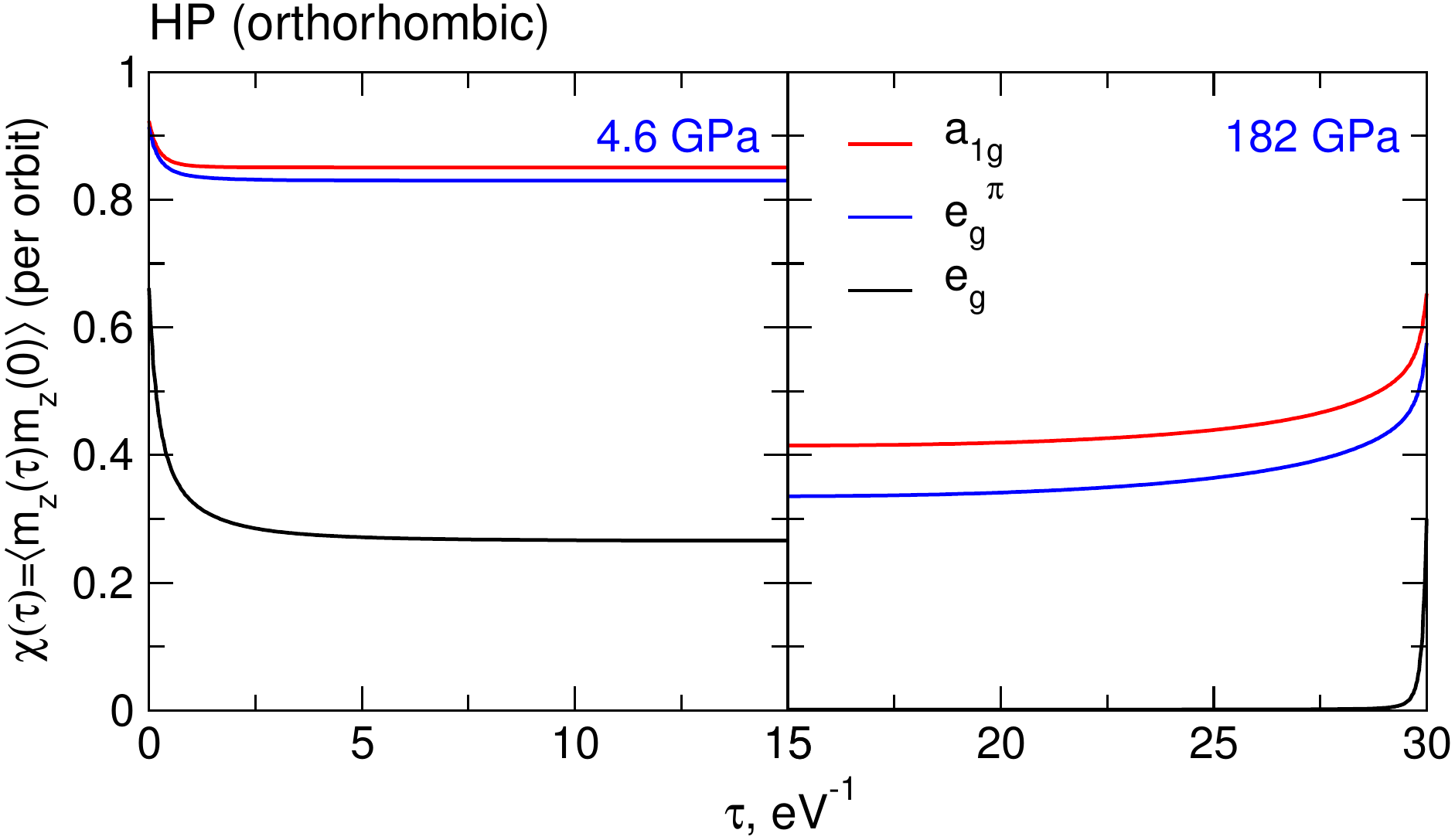}}
\caption{Orbitally-resolved local spin correlations functions $\chi(\tau)=\langle \hat{m}_z(\tau) \hat{m}_z(0) \rangle$ of the $Pbnm$ phase of CaFeO$_3$ calculated by DFT+DMFT for different pressures.
}
\label{Fig_5}
\end{figure}

The $Pbnm$ phase of CaFeO$_3$ is a strongly correlated metal, characterized by a Fermi-liquid-like behavior with a large orbital-dependent damping of $Im[\Sigma(\omega)]\sim0.52$ and 0.09 eV for the Fe $t_{2g}$ and $e_g$ quasiparticle states at the Fermi energy, at about 93~GPa. The latter indicates strong incoherence of the Fe $3d$ electronic states caused by proximity of the $3d$ statse to orbital-selective localization. We evaluate the quasiparticle mass enhancement $\frac{m^*}{m}=1-\partial Im\Sigma(\omega)/\partial \omega|_{\omega=0}$ using extrapolation of the self-energy $\Sigma(\omega)$ to $\omega=0$ eV, which gives a quantitative measure of the correlation strength. At 93~GPa we obtain $m^*/m \sim 5.0$ and 1.7 for the Fe $t_{2g}$ and $e_g$ bands, respectively. We also note that similarly to the insulating $P2_1/n$ phase the $Pbnm$ phase of CaFeO$_3$ exhibits a remarkable collapse of the local magnetic moments under pressure at about 43 GPa (which is almost equal to the critical pressure $p_c$ of the $P2_1/n$ to $Pbnm$ structural transition). 
This implies that the $P2_1/n$ to $Pbnm$ phase transition is accompanied by a collapse of the local magnetic moments of the Fe ions.

The MIT transition results in a significant redistribution of the spectral weight near the Fermi level. It leads to the formation of a quasiparticle peak originating from the Fe $t_{2g}$ states with the lower and upper Hubbard bands located at -1.2 eV and 1.5 eV, respectively (see the right panel of Fig.~\ref{Fig_3}). It gives a strong depopulation of the Fe $e_g$ orbital occupations. Thus, the Fe $e_g$ orbital occupations change from 0.38 (at ambient pressure in the $Pbnm$ phase) to 0.17 per spin-orbit under pressure increase to 93 GPa. 
On the other hand, it results in a gradual increase of the Fe $t_{2g}$ occupations, while the total Wannier Fe $3d$ occupation in the $Pbnm$ phase remains essentially unchanged with pressure, $\sim$4.73. We note that this value nearly equals to the average Wannier Fe $3d$ total occupancy of the Fe ions in the metallic $P2_1/n$ phase. 

In addition, our analysis of the atomic spin-state configurations and local spin susceptibilities shows that the phase transition is associate with a spin-state transition accompanied by a significant delocalization of the Fe $3d$ states (see Figs.~\ref{Fig_4} and \ref{Fig_5}). In particular, in the metallic $Pbnm$ phase at 93 GPa (i.e., above the MIT) the Fe ions adopt the IS state with a predominant weight of $\sim$56\%, with a strong admixture $\sim$34\% of the LS state.
Our analysis of fluctuating valence state of the Fe ions yields a predominant contribution of the $3d^5$ configuration, $\sim$47\%, with a considerable admixture of the $3d^4$ and $3d^6$ states of $\sim$34\% and 16\%, respectively. 
This implies that the electronic state of the Fe ions in the $Pbnm$ phase consists of heavily mixed $d^5\underline{L}$ and $d^4$ states, highlighting negative charge transfer due to strong covalency in CaFeO$_3$.
We therefore identify the metallic $Pbnm$ phase of CaFeO$_3$ as a negative charge-transfer material with a local Fe $3d^{5-\delta}\underline{L}^{1-\delta}$  electron configuration (with $\delta \ll 1$), implying the important role of Fe $3d$ to O $2p$ negative charge-transfer.

Overall, our results for the electronic structure, spin and charge state of Fe ions, and phase stability of CaFeO$_3$ under pressure agree well with available experimental studies. Our DFT+DMFT calculations give a microscopic understanding of the pressure-induced evolution of the electronic structure and lattice properties of CaFeO$_3$. It reveals the complex interplay between the electronic correlations and lattice effects in the vicinity of a pressure-induced MIT in CaFeO$_3$.
We note that our results are compatible with the scenario of bond disproportionation, which comprises the ``compressed'' Fe A $3d^{5-\delta}\underline{L}^{2-\delta}$ and ``expanded'' Fe B $3d^5$ sites with $\delta \ll 1$ alternating in the insulating $P2_1/n$ phase and a homogeneous Fe $3d^{5-\delta}\underline{L}^{1-\delta}$ state in the metallic $Pbnm$ phase.

%%%%%%%%%%%%%%%%%%%%%
\section{Conclusion}
%%%%%%%%%%%%%%%%%%%%%

In conclusion, we employed the DFT+DMFT computational approach to study the effects of electronic correlations on the electronic structure, spin and charge state of Fe ions, and structural phase stability of \emph{paramagnetic} CaFeO$_3$ across the pressure-induced insulator-to-metal phase transition. In agreement with experiments, the DFT+DMFT calculations show that at low pressure and temperature CaFeO$_3$ is a negative charge-transfer insulator characterized by strong localization of the Fe $3d$ electrons. It crystallizes in the monoclinic $P2_1/n$ crystal structure with a cooperative breathing mode distortion of the lattice.
While our analysis of the Fe $3d$ Wannier occupations and local moments for the insulating $P2_1/n$ phase give a robust charge disproportionation, a charge difference evaluated as an integral of the charge density around the structurally distinct Fe A and Fe B ions is rather weak, $\sim$0.04. This implies the importance of the Fe $3d$ and O $2p$ negative charge transfer and supports the formation of a bond-disproportionated state characterized by the Fe $3d^{5-\delta}\underline{L}^{2-\delta}$ and $3d^{5}$ valence configurations with $\delta \ll 1$ for the ``compressed'' Fe A and ``expanded'' Fe B sites, respectively, in the insulating phase of CaFeO$_3$. This behavior is consistent with a substantial Fe-O covalence and strong hybridization between the unoccupied Fe $e_g$ and O $2p$ states. 

In agreement with experiment, under critical pressure of $\sim$41~GPa CaFeO$_3$ is found to make the insulator-to-metal phase transition which is accompanied by a structural transformation into the orthorhombic $Pbnm$ phase. The phase transition is accompanied by suppression of the cooperative breathing mode distortion of the lattice and, hence, results in the melting of bond-disproportionation of the Fe ions. Our analysis suggests that the MIT transition is accompanied by orbital-dependent delocalization of the Fe $3d$ electrons and leads to a remarkable collapse of the local magnetic moments. Our results imply the crucial importance of the interplay of electronic correlations and structural effects to explain the electronic properties and structural phase equilibrium of CaFeO$_3$. We believe that this topic deserves further detailed theoretical and experimental considerations.

\begin{acknowledgments}

We acknowledge support by Russian Foundation for the Basic Research (Project No. 20-42-660027). The theoretical analysis of the electronic structure and DFT calculations were supported by the state assignment of Minobrnauki of Russia (theme ``Electron'' No. AAAA-A18-118020190098-5).

\end{acknowledgments}

\end{document}